# Early Outbreak Detection for Proactive Crisis Management Using Twitter Data: COVID-19 a Case Study in the US

Erfaneh Gharavi[a1], Neda Nazemi[a2], Faraz Dadgostari[a] *

[a] Engineering Systems and Environment, University of Virginia, VA, USA

**Abstract**
During a disease outbreak, timely non-medical interventions are critical in preventing the disease from growing into an epidemic and ultimately a pandemic. However, taking quick measures requires the capability to detect the early warning signs of the outbreak. This work collects Twitter posts surrounding the 2020 COVID-19 pandemic expressing the most common symptoms of COVID-19 including *cough* and *fever,* geolocated to the United States. Through examining the variation in Twitter activities at state level, we observed a temporal lag between the rises in the number of symptom reporting tweets and officially reported positive cases which varies between 5 to 19 days.

**Keywords**: Coronavirus, COVID-19, Social Media, Twitter Data, Symptoms, Outbreak Detection, Risk Management

1.  **Introduction**

*"Starting the new year off right with a cough and fever!"*
*"Starting the new year off right, sick as a dog with a high fever and a nasty cough. Craptastic."*
*"Starting 2020 with a fever and flu like symptoms is not how I pictured this decade starting"*
*"my ribs hurt when I cough so I don't want to cough but I have to cough I hate it here"*

These are only few examples of many Twitter messages (known as *tweets*) that people have posted in early 2020 in the United States, complaining about intense flu-like symptoms such as dry cough and fever, later on, were recognized as the most common symptoms of COVID-19.
SARS-CoV-2, the virus that causes COVID-19, is thought to have first transmitted from an animal host to humans in Wuhan, China in late 2019. On March 11, 2020, after the rapid increase of the cases outside China, the World Health Organization (WHO) eventually declared the COVID-19 as a pandemic (WHO 2020). As of April 25, it is officially reported that more than three million people are infected by this virus in 210 countries and territories around the world and 2 international conveyances (Worldometer 2020).
During a pandemic with a high infection rate, prompt mitigatory actions play a crucial role in decelerating the spread and preventing the new hotspots of the disease. Though, taking immediate actions requires the capability to detect the early warning signs of the outbreak and to characterize the dynamic of the spread in a near real-time fashion.

---

[1] E-mail address: eg8qe@virginia.edu
[2] E-mail address: nn2tf@virginia.edu
* corresponding author: fd4cd@virginia.edu

In the case of COVID-19 pandemic, delay in developing the test kits, the limited number of kits, complicated bureaucratic health care systems, and lack of transparency in data collection procedures are the major origins of postponement of effective preventive interventions and mitigatory (Washington post; Achrekar et al. 2011). Lai et. al. 2020, show if non-pharmaceutical interventions (NPIs) could have been conducted one week, two weeks, or three weeks earlier in China, cases would have been reduced by 66%, 86%, and 95%, respectively, together with significantly reducing the number of affected areas (Lai et al. 2020).

To fill this gap, Epidemic Intelligence (EI) is being used to explore alternative mostly informal sources of data to gather information regarding disease activity, early warning, and infectious disease outbreak (De Quincey and Kostkova 2010). Human activities and interactions on the web are one of these informal sources. For instance, Google Flu Trends exploits web search queries to estimate flu activity ("Google Flu Trends" 2012).

Social media content is another powerful tool that provides invaluable crowd-sourced near real-time data for sensing health trends.

Twitter is a microblogging service with around 330 million monthly active users that let users communicate through short messages (*tweets*) (Salman Aslam 2020). Twitter permits third parties to explore tweets and collect data about posters and their locations. It provides the opportunity to harness tweets data to detect early signs of outbreaks which can ultimately support decision-makers in taking more informed actions (Grover and Aujla 2014).

In this paper, we explore Twitter data right before and during COVID-19 pandemic across the United States at the state level, for the most common symptoms of COVID-19 including *cough* and *fever*. To offer a framework for outbreak early detection, the result of analysis on Twitter data are compared to the formal dataset provided by John Hopkins University which is openly available to the public for educational and academic research purposes[3].

The rest of this paper is organized as follows: Section 2 reviews the related literature that harness the Twitter data to analyze, detect and predict the outbreaks. In Section 3, we present our methodology for extracting relevant information from Twitter and preprocessing and analyzing the collected data. Elaborated results for six states are presented in Section 4. In Section 5, we will discuss the results, key findings and potential application, limitations and further steps of this study. Finally, we conclude in Section 6.

## 2. Related Work

Several studies have been conducted on the use of the Twitter data to explore the outbreak trends aiming to develop models for disease outbreak prediction. Achrekar et al. 2011 present a framework that monitors messages posted on Twitter with a mention of flu indicators to track and predict the emergence and spread of an influenza epidemic in a population (Achrekar et al. 2011).

---

[3] (https://github.com/CSSEGISandData/COVID-19)

Similarly, Chen et al. 2016, propose an approach to aggregate users' states in a geographical region for a better estimation of the flu trends (Chen et al. 2016). Smith et al. 2015, offer a method to distinguish between personal flu infection tweets versus general awareness tweets (i.e. expressing concern regarding flu outbreak) (Smith et al. 2015). The Twitter content during the 2009 H1N1 outbreak is analyzed by (Chew and Eysenbach 2010).

Besides flu, Tweeter data has been also leveraged to analyze other epidemics like Malaria, ZIKA virus, dengue, Ebola and so on. For instance, Masri et al. 2019, utilize tweet data in the United States to develop a prediction model of the number of ZIKA cases one week in advance while Maurice et al. 2019, offer a method to improve the Malaria surveillance system in Nigeria (Masri et al. 2019; Maurice et al. 2019). Current study addresses the gap of research on Twitter content during the most recent pandemic, COVID-19.

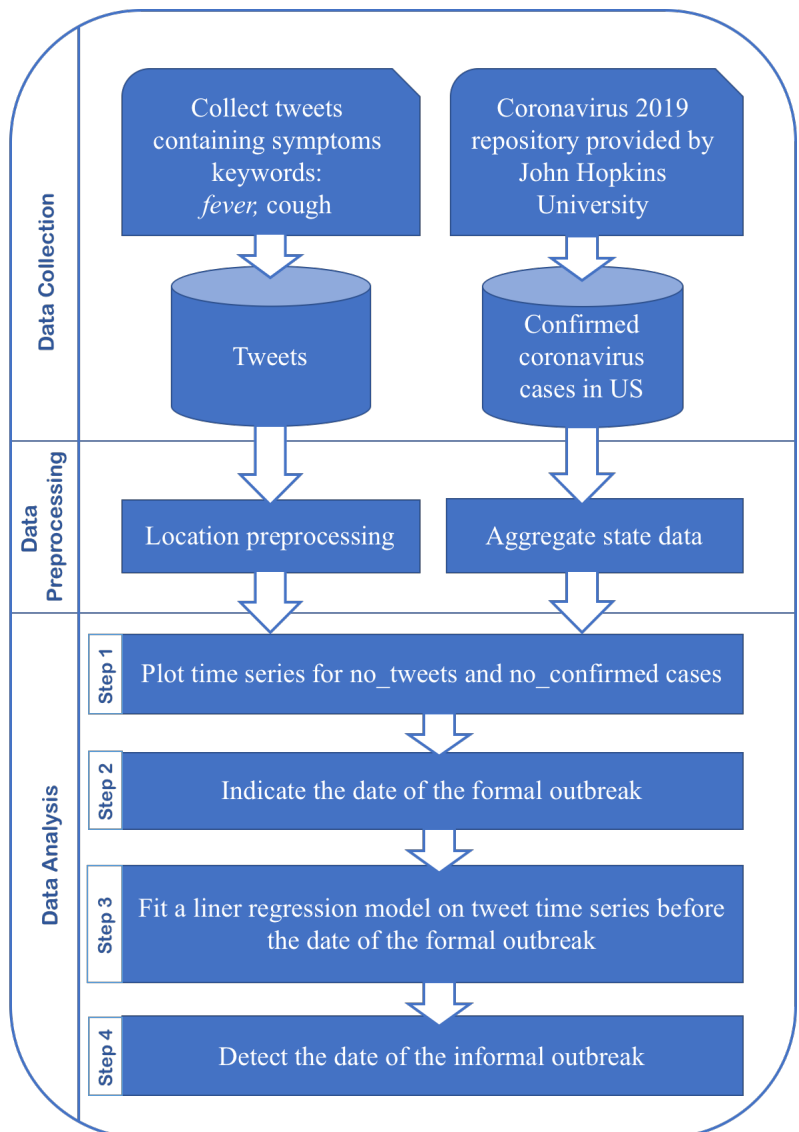

Figure 1: Proposed framework for early outbreak detection

## 3. Methodology

In this study, we propose a conceptual framework for investigating the temporal trends in the Twitter users' posts. This framework has three main modules including *Data Collection*, *Data Preprocessing* and *Data Analysis*. The framework schema is depicted in Figure 1. In the following section, we will further elaborate each module.

### 3.1 Data collection

We use GetOldTweets3[4] python package to retrieve historical tweets. By employing this package, the query can be restricted to get tweets containing determined keywords, during a particular time frame and within a specific region.

We collect tweets containing keywords *fever* or *cough*, as the main symptoms of coronavirus, from the beginning of September 2019 to April-16, 2020. The query limits the retrieved tweets to be within 1500-mile radius distance of Kansas City that covers all the states in America. Then, we use Twitter API (Application Programming Interfaces) to retrieve the corresponding tweets using the given IDs to access the precise geographical information of the user. The statistics of the number of tweets per state is shown in Figure 2. The data is available on our GitHub data repository[5].

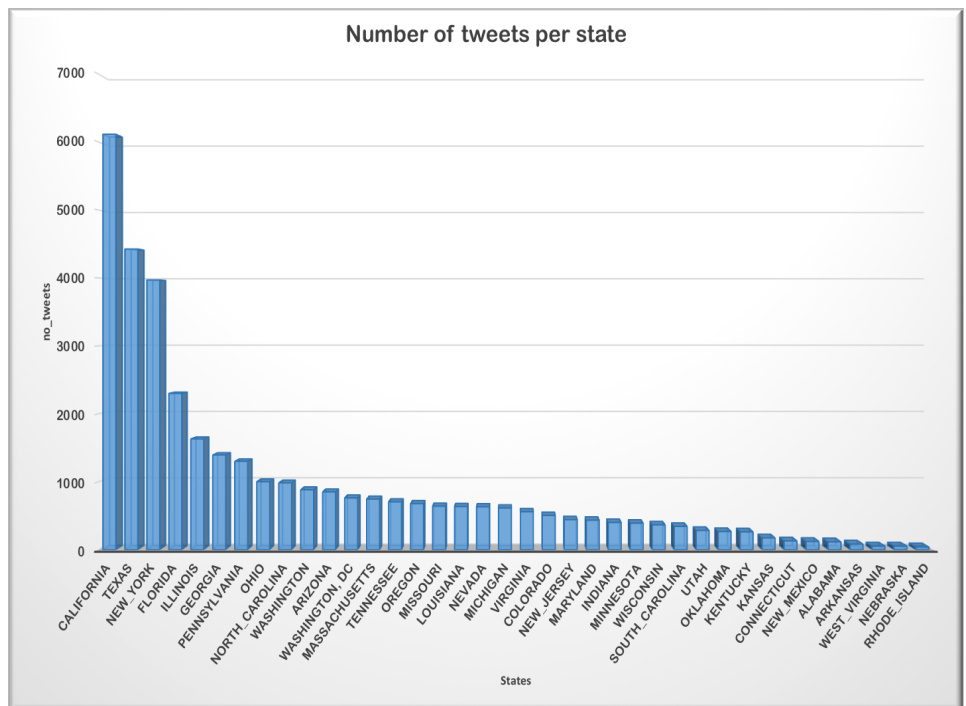

Figure 2: state-wise distribution of tweets containing symptom keywords

---

[4] https://pypi.org/project/GetOldTweets3/
[5] https://github.com/Erfanehgh/COVID19-Symptoms-tweetIDs.

To compare the results with the formal cases we use time series data of COVID -19 cases, reported by John Hopkins University. The data is reported from January 21 to April 16, 2020.

**3.2 Data preprocessing**

During the preprocessing, all the variations of the location name in Twitter data within a state are integrated into a unique token with the following format: "state_name, USA". It is the same naming format as in John Hopkin's dataset. Here are some examples:

**New_York, USA** $\Leftarrow$ {nyc, Rochester, NY, New York, USA, Staten Island, NY, Brooklyn, New York, USA, Bronx, NY, Manhattan, NY, Long Island, NY, Queens, NY, Buffalo, NY, New York City}
**Massachusetts, USA** $\Leftarrow$ {Boston, MA, Massachusetts, Boston}
**California, USA** $\Leftarrow$ {Fresno, CA, Southern California, Hesperia, CA, Los Angeles, California, Bakersfield, CA, Coachella Valley, CA, San Francisco, San Diego, Long Beach, CA, Los Angeles, CA}

The number of confirmed cases in John Hopkins dataset were reported separately for different counties within a state. The data has been aggregated over all counties in each state. For data preparation, a continuous time-series of the daily number of tweets and confirmed cases are calculated for each state.

**3.3 Data Analysis**

The data analysis steps are illustrated by Figure 3 on Colorado state data as an example.
**Step 1:** To compare the time series of the tweets containing COVID-19 symptoms and the number of confirmed cases, we plot the data from the beginning of December. We assign zero to all the dates before January 21, 2020 for the case data that was not available in the Johns Hopkins dataset.
**Step 2:** The date of the formal outbreak is defined as the date where the number of confirmed cases in a state exceeds 100 (Hartfield and Alizon 2013). We refer to this date as the beginning of the outbreak in a given state and show it with a vertical red line. As illustrated in Figure 3, the formal outbreak date at Colorado state was on March 14.
**Step 3:** For Colorado state and all other states**,** Tweet time series shows a linear growth trend from the beginning of December up to March 12th, followed by an exponential growth. To model the temporal trend, a regression-based estimator is fitted on the tweet data during this period which is represented by the black line.

**Step 4:** Finally, we detect the date of the *informal outbreak*, defined as the beginning of the exponential growth phase in tweets containing symptoms key words, by estimating the initial non-linearity on tweet time series. Vertical green line represents the *informal outbreak* on Figure 3.

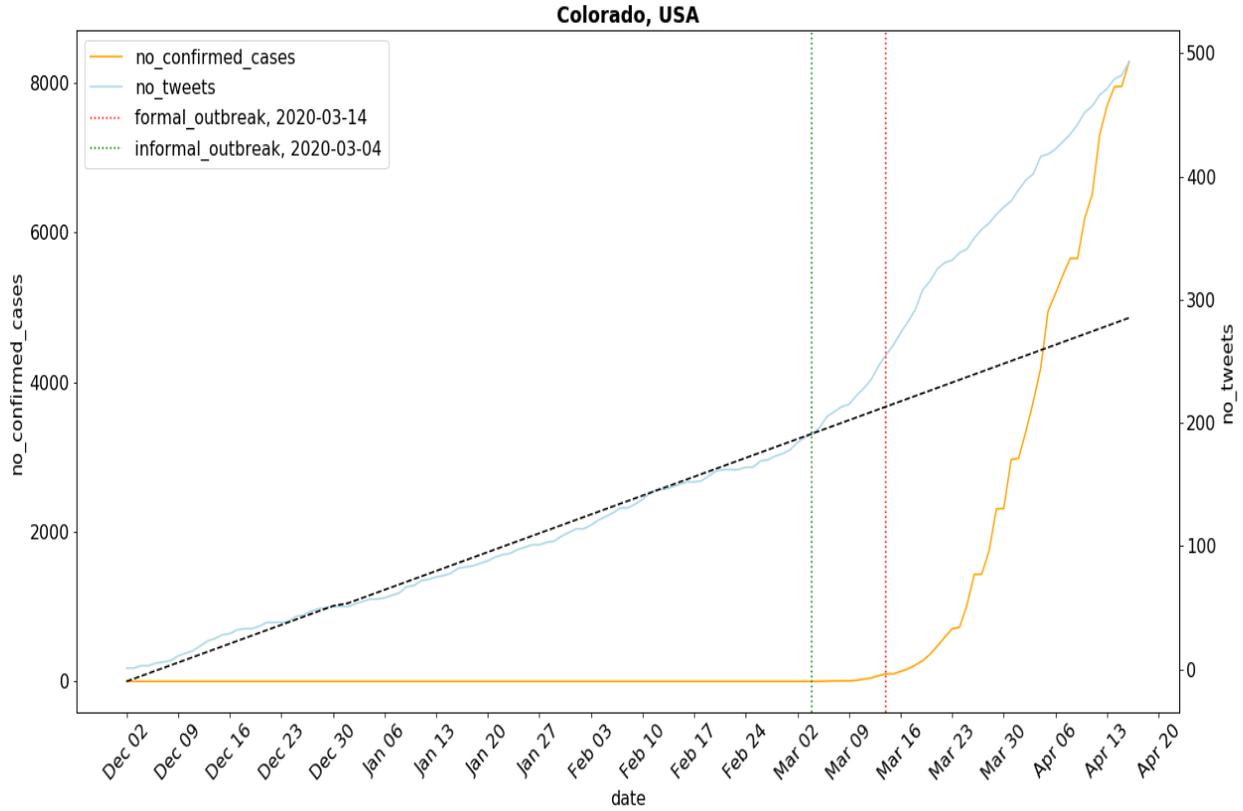

Figure 3: Comparison of the number of tweets (blue line) and number of confirmed cases (orange line) per day. The vertical red line represents the *formal outbreak*. Black line shows the trend of tweets before the formal outbreak, and vertical green line represents the *informal outbreak*.

1. Result

In this section, we present the results for six highly affected states. As explained in section 3.3, these plots (Figure 4) exhibit the number of tweets over time compared to the number of confirmed cases. The specified *formal outbreak* and the estimated *informal outbreak* are also shown for these states. This figure shows that there is a time lag between the estimated *informal outbreak* and *formal outbreak* which varies across the states. Table 1 summarizes the time lags observed for the six states shown in Figure 4. The longest and shortest time lags were detected as 19 and 7 days for Maryland and New York respectively. For most of the states the lag length is estimated around two weeks.

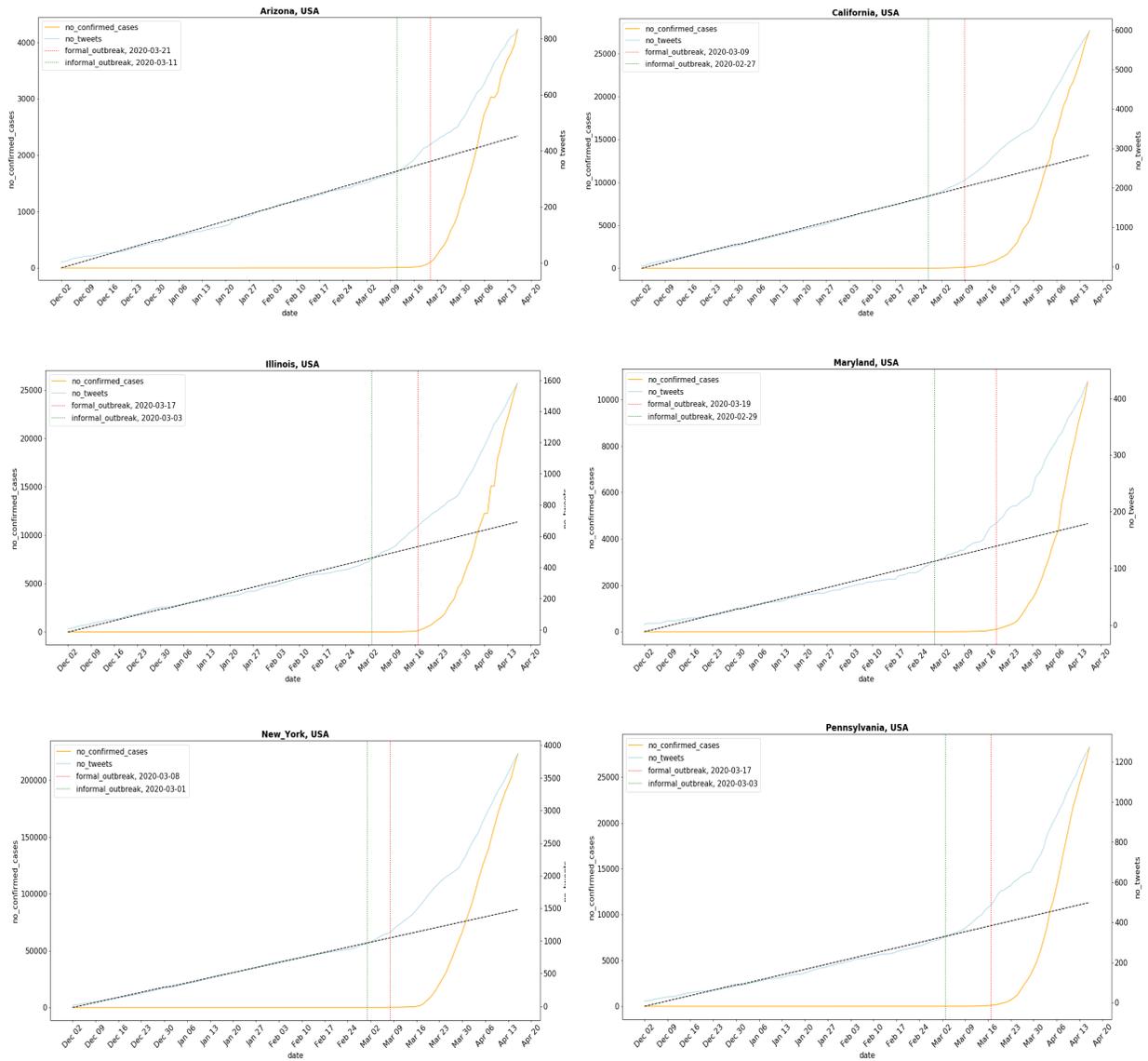

Figure 4: Results of the analysis for six states of Arizona, California, Illinois, Maryland, New York and Pennsylvania

Table 1: Time lag between estimated informal and indicated formal outbreaks

| State | Informal outbreak | Formal outbreak | Time Lag |
|---|---|---|---|
| Arizona | March-11-2020 | March-21-2020 | 10 |
| California | February-27-2020 | March-09-2020 | 11 |
| Illinois | March-03-2020 | March-17-2020 | 14 |
| Maryland | February-29-2020 | March-19-2020 | 19 |
| New York | March-01-2020 | March-08-2020 | 7 |
| Pennsylvania | March-03-2020 | March-17-2020 | 14 |

## 2. Discussion

Based on the epidemic models, usually in the early stages of an epidemic we expect an exponential growth trend in the number of the cases of disease (Martcheva 2015). However, it is difficult to monitor the growth trends of the infection in real-time and detect the outbreak without significant delay. This delay is often caused by the time-consuming and bureaucratic procedures of diagnosis including test development, test processing time, and reporting time (Rong et al. 2020). In this study we examined the possibility to fill this gap by detecting the early signs of an outbreak using Twitter content.

We collected tweets containing the common symptoms of COVID-19, right before and after the formal outbreak. A challenging issue in analyzing this data is to differentiate between general public concerns regarding the outbreak, and personal infection by COVID-19 to detect any anomaly in tweets' trend. To address this issue, Smith et al. 2015, use NLP to classify flu-related tweets into two categories of personal infection tweets that express an awareness of influenza. They show the temporal trends of these two categories are very different (Smith et al. 2015).

In this study, we assume that the general awareness tweets, prior to an outbreak in a given state, increases linearly, following the increase in the volume of the epidemic-related news of other countries or other states. On the other hand, we expect to observe exponential growth in the volume of the personal infection tweets, when an outbreak is happening in the given state, even prior to the official detection of the outbreak by formal medical procedures. Looking into the temporal trends in the Twitter data in 50 states of the US, prior to the official detection of the epidemic outbreak in any of the given states, as expected, we observe linear growth in the number of tweets following the news media reporting on the outbreak in china and later in western Europe. However, for each state there is a tipping point that happens before any official reports of the outbreak in those states where the growth trends change from linear to exponential, implying that the number of personal infection tweets not only dominate the general awareness tweets, but also define the growth behavior of the aggregate number of the tweets (implying that an outbreak is happening on top of a general awareness growth.)

In the case of COVID-19 pandemic, Lai et. al. 2020, show if non-pharmaceutical interventions were conducted one week, two weeks, or three weeks earlier in China, cases would have been reduced by 66%, 86%, and 95%, respectively, together with significantly reducing the number of affected areas (Lai et al. 2020). The observations of the current study, as a proof of concept, suggest that the behavioral patterns of an epidemic outbreak emerges in the temporal trends of the informal data streams like Twitter data, as an early sign of an outbreak in local level. In sum, this approach has potential to be used further as a decision support system to inform the policy makers deploying the intervention policies in a timely manner.

For future work, we suggest to validate the results of this study using a classifier to better differentiate the relevant from irrelevant tweets to exclude tweets containing 'baby fever' or 'cough cough'. Moreover, a model can be trained to monitor the fluctuation of symptom keyword usage and predict the pandemic in advance.

3. **Conclusion**

In this paper, we investigated the possibility of using Twitter content to detect and track COVID-19 outbreak in each state across the United States. We used a simple analysis of temporal trends of the relevant tweets. Our results have shown that the trend of *tweets* containing the common COVID-19 symptoms such as *cough* and *fever*, are highly correlated with the official CDC dataset. Although, a significant temporal lag, between 5 to 19 days, was observed between the exponential growth phase of *tweets* and the confirmed cases which could be related to inherent delays in the testing and diagnosis procedures. Therefore, we conclude that Twitter data provides a near real time assessment of an outbreak which can be utilized as an early warning system to increase the public awareness. It also can be used as a decision support system to inform policy maker in taking timelier mitigatory and preventive actions.